\documentclass[twocolumn,showpacs,preprintnumbers,amsmath,amssymb,nofootinbib,superscriptaddress,prc]{revtex4}
\usepackage{graphicx}
\usepackage{epstopdf}
\usepackage{dcolumn}
\usepackage{bm}
\usepackage[latin2]{inputenc} 
\usepackage{threeparttable}
\usepackage{longtable}
\usepackage{color}
\usepackage{soul}
\usepackage{enumitem}

\begin{document}

\preprint{}

\title{Quenching preheating by light fields}

\author{Olga Czerwi\'nska}
\affiliation{Institute of Theoretical Physics, Faculty of Physics, University of Warsaw ul. Pasteura 5, 02-093 Warsaw, Poland}
\author{Seishi Enomoto}
\affiliation{Institute of Theoretical Physics, Faculty of Physics, University of Warsaw ul. Pasteura 5, 02-093 Warsaw, Poland}
\affiliation{University of Florida, Department of Physics, P.O. Box 118440, Gainesville, FL 32611-8440}
\author{Zygmunt Lalak}
\affiliation{Institute of Theoretical Physics, Faculty of Physics, University of Warsaw ul. Pasteura 5, 02-093 Warsaw, Poland}

\date{\today}

\begin{abstract}

In this paper we investigate the role of additional light fields not directly coupled to the background during preheating. We extend our previous study that proved that the production of particles associated with such fields can be abundant due to quantum corrections, even for the massless states. We also obtain the expression for the occupation number operator in terms of interacting fields which includes the non-linear effects important for non-perturbative particle production. We show that adding too many light degrees of freedom without direct interactions with the background might attenuate or even quench preheating as the result of back-reaction effects and quantum corrections.

\end{abstract}

\pacs{98.80.-k, 98.80.Cq}

\maketitle

\section{Introduction}

Post-inflationary particle production is a very complex stage in the evolution of the universe that mixes perturbative and non-perturbative processes \cite{Kofman:1994rk, Kofman:1997yn, Traschen:1990sw, Dolgov:1989us, Kofman:2004yc}. Usually it is divided into two main stages:
\begin{enumerate}[label=\alph*)]
\item preheating - when exponentially and non-perturbatively produced states typically correspond to the fields directly interacting with the inflaton, they do affect the mass term of the inflaton through back-reaction effects
\item reheating (thermalization) - when the inflaton decays perturbatively and produced particles end up in thermal equilibrium with a well-defined temperature.
\end{enumerate} 

Interesting is the question about the impact of the additional fields, especially light ones, on preheating. Their presence in the theory during \cite{Kobayashi:2010fm, Matsuda:2012kk, Kohri:2014jma} and after inflation is important for multi-field inflation models and for curvaton scenarios \cite{Enqvist:2001zp, Lyth:2001nq, Moroi:2001ct}. For recent reviews of post-inflationary particle production see \cite{Allahverdi:2010xz} or \cite{Amin:2014eta}. 

In our previous study \cite{Enomoto:2014cna} we showed that light fields which are not coupled directly to the background can be produced due to quantum corrections and their abundance can be sizeable, even for the massless case. In this paper we want to develop these results addressing the problem of additional light degrees of freedom and avoiding at the same time the infinite growth resulting from the approximation used previously. The crucial difference between present considerations at that of \cite{Enomoto:2014cna} lies in the fact that presently the inflaton is massive.

The outline of the paper is as follows. In Section \ref{sec_process} we develop the formalism necessary to describe the creation of particles in the presence of interactions, with and without time-varying vacuum expectation value (vev) of the considered field. In Section \ref{sec_ex2} we apply our formalism to a number of well-motivated cosmological scenarios, including a sector of very light fields. In Section \ref{sec_disc} we compare our results with the earlier ones \cite{Enomoto:2014cna}, discuss the role of different parameters in the theory, summarize the paper and conclude.

\section{Particle production in terms of interacting fields \label{sec_process}}

Usually the occupation number operator of produced particles is defined in terms of the creation and annihilation operators as $N_{\mathbf{k}}\equiv a_{\mathbf{k}}^\dagger a_{\mathbf{k}}$.  This definition assumes that produced states can be treated as free fields which means that their equations of motion are linear. However, in general fields associated with the produced particles interact with other fields which spoils linearity and results in the non-perturbative production. In that case it is not clear how to define the number operator properly. In this section we address this issue and describe particle number using the theory of interacting fields which takes into account the non-linear effects. To compare these results with a simpler theory of a free field with time-dependent mass term see Appendix A.

For simplicity let us consider a real scalar field $\phi$ with the Lagrangian of the form:
\begin{equation}
 \mathcal{L} = \frac{1}{2} (\partial \phi)^2 - \frac{1}{2}m_0^2 \phi^2 - V[\phi,({\text{other fields}})], \label{lagr}
\end{equation}
where $m_0$ is a bare mass of $\phi$ and $V$ is a general potential. Then equation of motion reads: 
\begin{eqnarray}
 0 = (\partial^2+m_0^2)\phi+\frac{\partial V}{\partial \phi} = (\partial^2+M^2)\phi+J, \label{eom}
\end{eqnarray}
where $M$ is a physical mass that can depend on time \footnote{In general physical mass can depend not only on time but also on space coordinates. For simplicity we consider only the time-dependent case as it is more common in cosmological considerations.} and should be a c-number, and
\begin{equation}
 J \equiv (m_0^2-M^2)\phi+\frac{\partial V}{\partial \phi}
\end{equation}
is a source term that can be an operator.

Formal solution of (\ref{eom}) can be presented in a form of the Yang-Feldman equation
\begin{equation}
  \phi(x)=\phi^{(t)}(x)-\int_t^{x^0}d^4y \: i[\phi^{(t)}(x), \phi^{(t)}(y)]J(y), \label{eq:YF_real_scalar}
\end{equation}
where the first term describes an asymptotic field defined at $x^0=t$ which satisfies the free field equation of motion
\begin{eqnarray}
 0 &=& (\partial^2+M^2)\phi^{(t)}.
\end{eqnarray}
In case that $\phi$ does not have a vev,  the $\phi^{(t)}$ can be decomposed into modes
\begin{equation}
 \phi^{(t)}(x) = \int\frac{d^3k}{(2\pi)^3}e^{i\mathbf{k\cdot x}}\left(
  \phi_k^{(t)}a_{\mathbf{k}}^{(t)} + \phi_k^{(t) *}a_{\mathbf{-k}}^{{(t)}\dagger} \right)
 \label{eq:asym_real_scalar}
\end{equation}
fulfilling harmonic oscillator equation
\begin{equation}
 0 = \ddot{\phi}_k^{(t)} + \omega_k^2 \phi_k^{(t)}
\end{equation}
with $k\equiv |\mathbf{k}|$, $\omega_k\equiv\sqrt{k^2+M^2}$ and obeys the inner product \footnote{We use the following definition of the inner product: 
\begin{equation}
(A,B)\equiv i(A^\dagger \dot{B}-\dot{A}^\dagger B). \nonumber 
\end{equation}} relation: $\left(\phi_k^{(t)},\phi_k^{(t)}\right)=1$.

From (\ref{eq:YF_real_scalar}) also the relation between two asymptotic fields defined at different times $x^0=t$ and $x^0=t^{\rm in}$ can be derived 
\begin{equation}
 \phi^{(t)}(x) = \phi^{\rm in}(x)-\int_{t^{\rm in}}^{t}d^4y \: i[\phi^{\rm in}(x), \phi^{\rm in}(y)]J(y),\label{eq:relation_t_in}
\end{equation}
where we denoted $\phi^{\text{in}} (x) \equiv \phi^{(t^{\text{in}})} (x) $. Evaluating the inner product of the above equation with $\phi_k^{(t)}$: $\left( \phi^{(t)}, \phi_k^{(t)} \right)$, we can obtain the Bogoliubov transformation for annihilation operators \cite{Enomoto:2014cna}
\begin{equation}
 a_{\mathbf{k}}^{(t)}=\alpha_ka_{\mathbf{k}}^{\rm in}+\beta_ka_{-\mathbf{k}}^{{\rm in}\dagger}
   - \int_{t^{\rm in}}^td^4y \: i[\alpha_k a_{\mathbf{k}}^{\rm in}+\beta_ka_{\mathbf{-k}}^{{\rm in}\dagger},
  \phi^{\rm }(y)]J(y), \label{eq:BTA_real_scalar}
\end{equation}
where
\begin{equation}
 \alpha_k=\alpha_k(t,t^{\rm in})\equiv(\phi_k^{(t)},\phi_k^{\rm in}), \quad \beta_k=\beta_k(t,t^{\rm in})\equiv(\phi_k^{(t)},\phi_k^{{\rm in}*})
\end{equation}
and the normalization condition reads
\begin{equation}
|\alpha_k|^2-|\beta_k|^2=1.
\end{equation}
These relations are equivalent to the Bogoliubov transformation for the wave function
\begin{equation}
 \phi_k^{(t)}=\alpha_k^*\phi_k^{\rm as}-\beta_k^*\phi_k^{{\rm as}*},
\end{equation}
where $\phi_k^{\rm as}$ denotes the asymptotic value of the mode $\phi_k$, $\phi_k^{\rm as} = \phi_k^{\rm in}, \phi_k^{\rm out} $.

Lagrangian (\ref{lagr}) corresponds to the Hamiltonian
\begin{equation}
H = \int d^3x\left[\frac{1}{2}\dot{\phi}^2
  +\frac{1}{2}(\nabla\phi)^2+\frac{1}{2}m_0^2\phi^2+V\right].
\end{equation}
Substituting (\ref{eq:relation_t_in}) into the above results in a quite complicated expression for the Hamiltonian which can be simplified by choosing the Bogoliubov coefficients of the form
\begin{equation}
 |\alpha_k|^2=\frac{\Omega_k^{\rm in}}{2\omega_k}+\frac{1}{2}, \hspace{0.1cm}
 |\beta_k|^2=\frac{\Omega_k^{\rm in}}{2\omega_k}-\frac{1}{2}, \hspace{0.1cm}
 {\rm Arg}(\alpha_k \beta_k^*) = {\rm Arg} \: \Lambda_k^{\rm in}
 \label{eq:BCs_concrete_form}
\end{equation}
where \footnote{$\Omega_k^{\rm in}$ and $\Lambda_k^{\rm in}$ are constrained by the relation: 
\begin{equation}
\nonumber |\Omega_k^{\rm in}|^2-|\Lambda_k^{\rm in}|^2=\omega_k^2.
\end{equation} 
}
\begin{equation}
 \Omega_k^{\rm in}\equiv|\dot{\phi}_k^{\rm in}|^2+\omega_k^2|\phi_k^{\rm in}|^2,
  \quad \Lambda_k^{\rm in}\equiv(\dot{\phi}_k^{\rm in})^2+\omega_k^2(\phi_k^{\rm in})^2. \label{eq:Omega_Lambda}
\end{equation}
Then the Hamiltonian with diagonalized kinetic terms reads
\begin{eqnarray}
& H = \int\frac{d^3k}{(2\pi)^3}\:\omega_k\left(
  a_{\mathbf{k}}^{(t)\dagger}a_{\mathbf{k}}^{(t)}+\frac{1}{2} (2\pi)^3\delta^3(\mathbf{k}=0)\right) + \\
  & \nonumber  + \int d^3x \left[\frac{1}{2}(m_0^2-M^2)\phi^2+V\right], \label{eq:Hamiltonian_interacted2}
\end{eqnarray}
which indicates that the operator
\begin{equation}
 N_{\mathbf{k}}(t)\equiv a_{\mathbf{k}}^{(t)\dagger}a_{\mathbf{k}}^{(t)} \label{eq:number_op_real_scalar}
\end{equation}
really plays the role of the occupation number. This is because in the system which has a potential energy particle number $N$ would be described classically as
\begin{equation}
N = \frac{H - V_{\rm eff} - V_0}{E},
\end{equation}
where $H$ is the total Hamiltonian, $V_{\rm eff}$ an effective potential, $V_0$ a zero-point energy and $E$ is an one-particle energy. Therefore particle number is just the kinetic energy of the system divided by the one-particle energy.

Substituting (\ref{eq:BTA_real_scalar}) into (\ref{eq:number_op_real_scalar}) and using (\ref{eq:BCs_concrete_form}), (\ref{eq:Omega_Lambda}), we can finally obtain the expression for the occupation operator in terms of the interacting fields as
\begin{equation}
N_{\textbf{k}} (t) = \frac{1}{2} \left[ N^{+}_{\textbf{k}} (t) + N^{-}_{\textbf{k}} (t)\right]
\end{equation}
with\footnote{The zero point term can be regarded as the volume of the system because
\begin{equation}
 (2\pi)^3\delta^3(\mathbf{k}=0)=\int d^3x e^{i\mathbf{k\cdot x}}|_{\mathbf{k}=0}=\int d^3x = V.
\end{equation}
Therefore, we can also find distribution operators:
\begin{eqnarray}
& n^{+}_{\textbf{k}} = \frac{N^{+}_{\textbf{k}}}{V} = \frac{1}{\omega_k} \left( \frac{1}{V}\dot{\hat{\phi}}^{\dagger}_{\textbf{k}} \dot{\hat{\phi}}_{\textbf{k}} + \omega_k^2 \cdot\frac{1}{V}\hat{\phi}_{\textbf{k}}^{\dagger} \hat{\phi}_{\textbf{k}} \right) - 1,\\
& n^{-}_{\textbf{k}} = \frac{N^{-}_{\textbf{k}}}{V} = i \left( \frac{1}{V}\hat{\phi}^{\dagger}_{\textbf{k}} \dot{\hat{\phi}}_{\textbf{k}} - \frac{1}{V}\dot{\hat{\phi}}_{\textbf{k}}^{\dagger} \hat{\phi}_{\textbf{k}} \right) +1.
\end{eqnarray}
}
\begin{eqnarray}
& N^{+}_{\textbf{k}} (t) = \frac{1}{\omega_k} \left( \dot{\hat{\phi}}^{\dagger}_{\textbf{k}} \dot{\hat{\phi}}_{\textbf{k}} + \omega_k^2 \hat{\phi}_{\textbf{k}}^{\dagger} \hat{\phi}_{\textbf{k}} \right) - (2\pi)^3\delta^3(\mathbf{k}=0)
, \label{eq:n+}\\
& N^{-}_{\textbf{k}} (t) = i \left( \hat{\phi}^{\dagger}_{\textbf{k}} \dot{\hat{\phi}}_{\textbf{k}} - \dot{\hat{\phi}}_{\textbf{k}}^{\dagger} \hat{\phi}_{\textbf{k}} \right) + (2\pi)^3\delta^3(\mathbf{k}=0)
, \label{eq:n-}
\end{eqnarray}
where we defined the Fourier transformation as
\begin{equation}
\hat{\phi}_{\textbf{k}} (t) \equiv \int d^3 x e^{-i \textbf{k} \cdot \textbf{x}} \phi(t, \textbf{x}). \label{eq:vev}
\end{equation}
Since $N^{\pm}_{\textbf{k}} (t) = a_{\textbf{k}}^{(t) \dagger} a_{\textbf{k}}^{(t)} \pm a_{-\textbf{k}}^{(t) \dagger} a_{-\textbf{k}}^{(t)} $, $N^{+}_{\textbf{k}} $ denotes a total and $N^{-}_{\textbf{k}} $ a net number of particles with momentum between \textbf{k} and -\textbf{k}. In the case of a complex scalar $\phi$ this expression changes to $ N^{\pm}_{\textbf{k}} (t)~=~a_{\textbf{k}}^{(t) \dagger} a_{\textbf{k}}^{(t)} \pm b_{-\textbf{k}}^{(t) \dagger} b_{-\textbf{k}}^{(t)} $, where $ b_{\textbf{k}}^{(t)}$ is an annihilation operator for anti-state, but (\ref{eq:n+}) and (\ref{eq:n-}) still hold. In such a case $\int \frac{ d^3 k}{(2 \pi)^3} N^-_{\textbf{k}} $ corresponds to the $U(1)$ Noether charge.

In the case where $\phi$ has a non-vanishing vev $\left<\phi\right>\equiv\left<0^{\rm in}\right|\phi\left|0^{\rm in}\right>$, we just have to replace $\phi \rightarrow \phi - \left<\phi\right>$ in (\ref{eq:vev}) to obtain the proper expression for the occupation number.

\section{Numerical results for multi-scalar systems \label{sec_ex2}}

To obtain numerical results for some specific models we follow the procedure described in Section \ref{sec_process}. We are especially interested in time-evolution of particle number density for each considered species:
\begin{equation}
n(t) = \int \frac{d^3 k}{(2 \pi)^3} \frac{\langle N_{\mathbf{k}} \rangle}{V},
\end{equation}
where $V$ is the volume of the system, and in time-dependence of the background (inflaton). We consider a time range and starting from the initial state we solve equations of motion for all the species and calculate their number density. Then we move to a slightly later time and repeat the procedure taking into account the back-reaction of previously produced states on the evolution of the background (given by the induced potential coming from non-zero energy density) and all the species.

Before we present numerical results for specific models, let us focus on a subtlety in calculation of the particle number with a general Lagrangian (\ref{lagr}). In order to describe the time evolution of distributions $n_{\textbf{k}} = \langle N_{\textbf{k}} \rangle/V$ for each type of produced states we need to determine the time evolution of bilinear products of field operators $ \langle \dot{\hat{\phi}}_{\textbf{k}}^{\dagger} \dot{\hat{\phi}}_{\textbf{k}} \rangle $, $ \langle \hat{\phi}_{\textbf{k}}^{\dagger} \hat{\phi}_{\textbf{k}} \rangle $ and $ \langle \hat{\phi}_{\textbf{k}}^{\dagger} \dot{\hat{\phi}}_{\textbf{k}} \rangle $. Equations of motion for these operators can be derived by calculating their time derivatives and using (\ref{eom})as
\begin{eqnarray}
 \langle \hat{\phi}_{\textbf{k}}^{\dagger} \hat{\phi}_{\textbf{k}} \rangle^{\cdot}
  &=& \langle \dot{\hat{\phi}}_{\textbf{k}}^{\dagger}\hat{\phi}_{\textbf{k}} \rangle + \langle \hat{\phi}_{\textbf{k}}^{\dagger} \dot{\hat{\phi}}_{\textbf{k}} \rangle \label{eq:pp_d}\\
 \langle \hat{\phi}_{\textbf{k}}^{\dagger} \dot{\hat{\phi}}_{\textbf{k}} \rangle^{\cdot}
  &=& \langle \dot{\hat{\phi}}_{\textbf{k}}^{\dagger} \dot{\hat{\phi}}_{\textbf{k}} \rangle + \langle \hat{\phi}_{\textbf{k}}^{\dagger} \ddot{\hat{\phi}}_{\textbf{k}} \rangle \nonumber \\
  &=& \langle \dot{\hat{\phi}}_{\textbf{k}}^{\dagger} \dot{\hat{\phi}}_{\textbf{k}} \rangle - \omega_k^2 \langle \hat{\phi}_{\textbf{k}}^{\dagger} \hat{\phi}_{\textbf{k}} \rangle - \langle \hat{\phi}_{\textbf{k}}^{\dagger} \hat{J}_{\textbf{k}} \rangle \label{eq:pd_d}\\
 \langle \dot{\hat{\phi}}_{\textbf{k}}^{\dagger} \dot{\hat{\phi}}_{\textbf{k}} \rangle^{\cdot}
  &=& \langle \ddot{\hat{\phi}}_{\textbf{k}}^{\dagger}\dot{\hat{\phi}}_{\textbf{k}} \rangle + \langle \dot{\hat{\phi}}_{\textbf{k}}^{\dagger} \ddot{\hat{\phi}}_{\textbf{k}} \rangle \nonumber \\
  &=& -\omega_k^2(\langle \dot{\hat{\phi}}_{\textbf{k}}^{\dagger} \hat{\phi}_{\textbf{k}} \rangle + \langle \hat{\phi}_{\textbf{k}}^{\dagger}\dot{\hat{\phi}}_{\textbf{k}} \rangle)  -\langle \dot{\hat{\phi}}_{\textbf{k}}^{\dagger} \hat{J}_{\textbf{k}} \rangle - \langle \hat{J}_{\textbf{k}}^{\dagger}\dot{\hat{\phi}}_{\textbf{k}} \rangle \nonumber \label{eq:dd_d}\\
\end{eqnarray}

where
\begin{equation}
\hat{J}_{\textbf{k}} \equiv \int d^3 x e^{- \textbf{k} \cdot \textbf{x}} J(t, \textbf{x}).
\end{equation}
Physical mass of $\phi$ is determined by the relation:
\begin{eqnarray}
& 0 = \langle \hat{\phi}_{\textbf{k}}^{\dagger} \hat{J}_{\textbf{k}} \rangle = (m^2 - M^2) \langle \hat{\phi}_{\textbf{k}}^\dagger \hat{\phi}_{\textbf{k}} \rangle \\
& \nonumber + \int d^3 x e^{-i \mathbf{k\cdot x}} \left \langle \hat{\phi}_{\textbf{k}}^\dagger \frac{dV(x)}{ d\phi} \right \rangle
\end{eqnarray}
to remove the infinite part of the mass correction.

\subsection{Two scalar system}

At first we apply our formalism to the simple theory consisting of two scalar fields
\begin{equation}
 \mathcal{L} = \frac{1}{2}(\partial \phi)^2 + \frac{1}{2}(\partial \chi)^2
   - \frac{1}{2}m_\phi^2 \phi^2 - \frac{1}{2}m_\chi^2 \chi^2 - \frac{1}{4}g^2\phi^2\chi^2. \label{lagr2}
\end{equation}
We assume that it is the field $\phi$ that has time-varying vev and plays the role of inflaton, $ \left<0^{\rm in}\right|\phi\left|0^{\rm in}\right>=\left<\phi(t)\right>$, while $\chi$ is another scalar field with vanishing vev that can be, for instance. a mediator field between the inflaton and the Standard Model. We also assume $m_\phi \gg m_\chi$. The details of the calculation in this system can be found in Appendix B.

Asymptotically, when quantum effects can be neglected, we can choose a vacuum solution for (\ref{lagr2}) of the form
\begin{equation}
\langle \phi \rangle = \phi_0 \cos (m_{\phi} (t-t_0)),
\end{equation}
where $\phi_0$ denotes the initial amplitude of the oscillations, $\langle \phi \rangle (t=t_0) = \phi_0$. When this trajectory crosses the non-adiabatic area for $\chi$:  $ |\langle \phi \rangle| < \sqrt{m_{\phi} |\phi_0|/g}$, the mass of $\chi$ becomes very small and kinetic energy of the background field $\phi$ is transferred to the field $\chi$. This results in the creation of $\chi$ particles with the distribution \citep{Kofman:2004yc}
\begin{equation}
n_k = e^{-\pi \frac{k^2}{g m_{\phi} |\phi_0|}},
\end{equation}
where $k$ is a momentum of a $\chi$ particle. Once $\chi$ particles are produced and trajectory of $\langle \phi \rangle$ goes away from the non-adiabatic region, the energy density of $\chi$ particles can be represented as 
\begin{equation}
\rho_{\chi} \sim g |\langle \phi \rangle| \int \frac{d^3 k}{(2 \pi)^3} n_k,
\end{equation}
which corresponds to the linear potential acting on $\langle \phi \rangle$ describing the backreaction effects. Then trajectory of $\langle \phi \rangle$ goes back towards the origin and $\chi$ particles can be produced again both due to the  oscillatory behaviour of $\langle \phi \rangle $ and backreaction.

\begin{figure}[h!]
\begin{center}
\scalebox{0.35}{\includegraphics[width=1.1\textwidth]{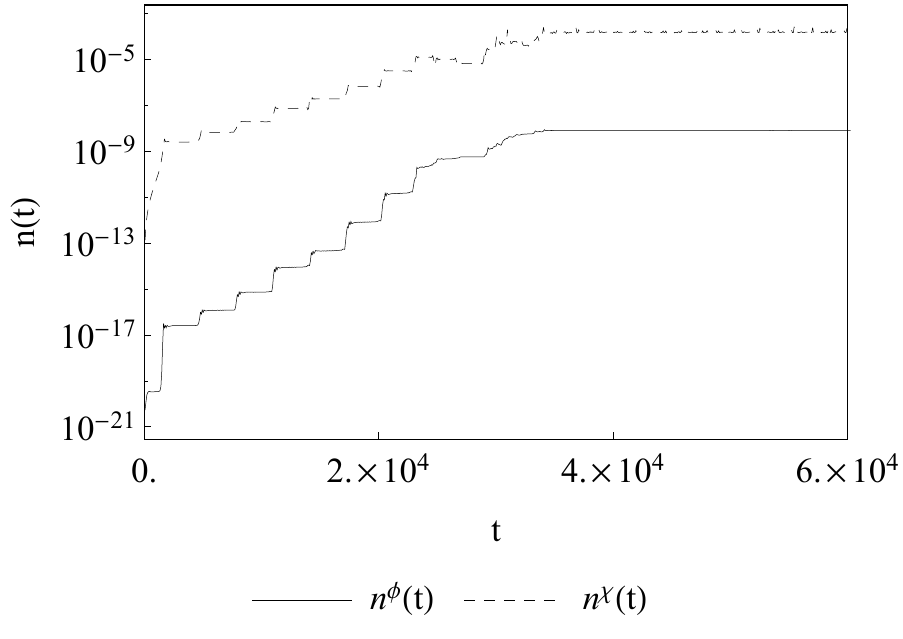}}
\caption{Time evolution of number density of produced states for $g=0.1$, $m_\phi=0.001 M$, $\phi~(t=~0)=M$, $\dot{\phi}(t=0)=0$ in two scalar system. Scale $M \sim 0.04 M_{PL}$, where $M_{PL}$ denotes the Planck mass $M_{PL} \sim 1.22 \cdot 10^{19}$ GeV, is chosen to be close to the unification scale and allows us to stay in agreement with the observational data.} \label{fig:two}
\end{center}
\end{figure}

In the Figure \ref{fig:two} we show an example of the numerical results for the Lagrangian (\ref{lagr2}). According to \cite{Kofman:2004yc} the first production of $\chi$ particles results in the number density
\begin{equation}
 n^{(1)}_\chi \sim \frac{(gm_\phi\langle\phi(0)\rangle)^{3/2}}{(2\pi)^3}\sim4\times10^{-9},
\end{equation}
which is consistent with our numerical results. On the other hand, it is difficult to obtain the analytic results for indirect production products, like $\tilde{\phi}\equiv\phi-\langle\phi\rangle$. But one can see in the Figure \ref{fig:two} that for the considered Lagrangian energy transfer from the background to $\tilde{\phi} $ and the production of particles associated with the inflaton is small for generic choices of parameters. Therefore in this system it is a good approximation to neglect the quantum part of $\tilde{\phi}$ and the production of its fluctuation.

\begin{figure}[h!]
\begin{center}
\scalebox{0.35}{\includegraphics[width=1.1\textwidth]{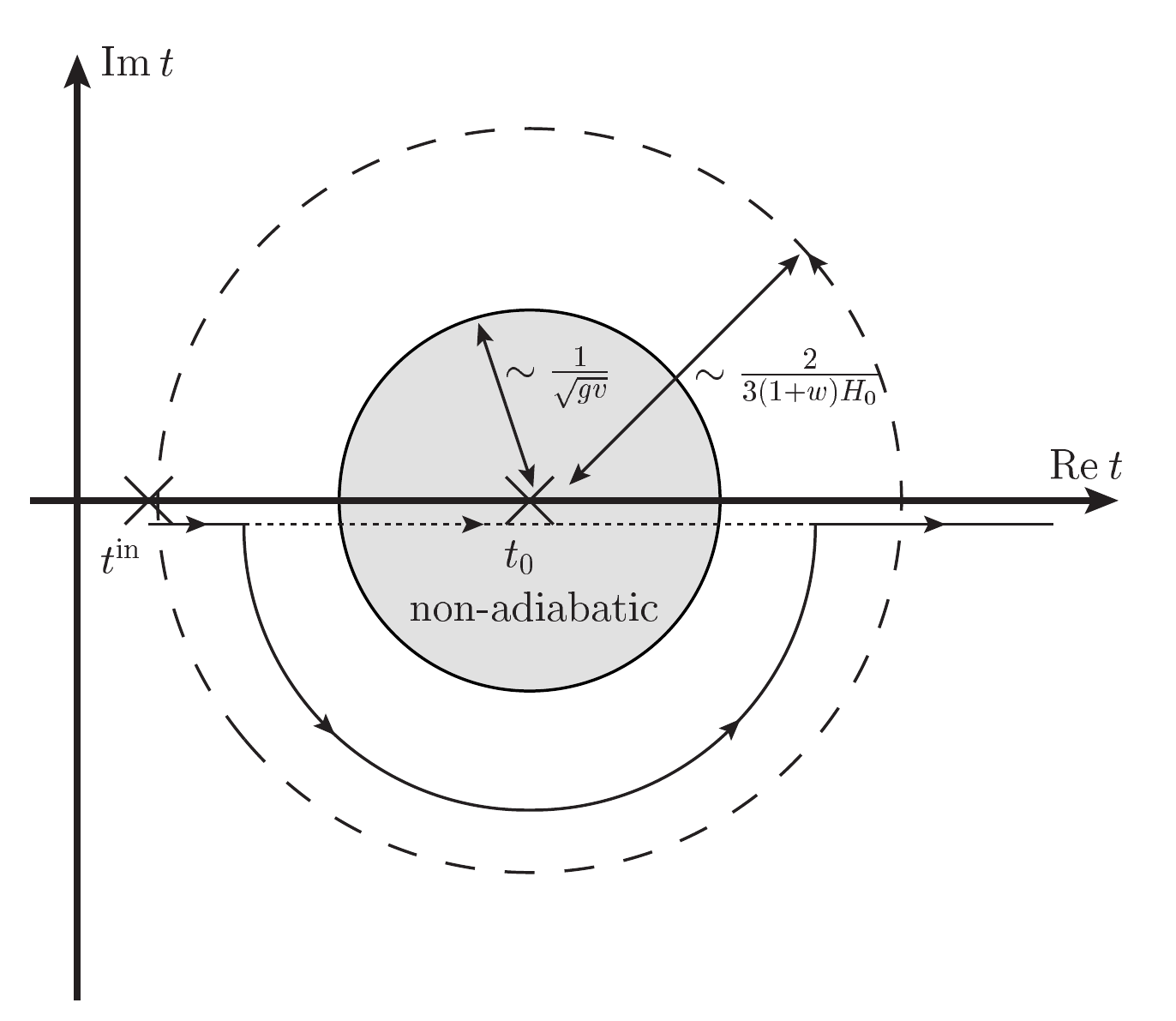}}
\caption{ Time spent by the trajectory in the adiabatic region in comparison with the Hubble time.}
\label{fig:gravity}
\end{center}
\end{figure}

In our considerations we neglect the expansion of the universe which is valid assuming that the mean time the trajectory spends in the non-adiabatic region is smaller than the Hubble time, see Figure \ref{fig:gravity}. This means that:
\begin{equation}
\frac{1}{\sqrt{gv}} < \frac{2}{3 H (w+1)},
\end{equation}
where $H$ is a Hubble parameter and $w = \frac{p}{\rho}$ is a barotropic parameter describing the content of the universe. For the scalar field domination phase this means that $\sqrt{gv} > 3 H $, for matter domination: $ \sqrt{gv} > \frac{3}{2} H$, while for radiation domination: $\sqrt{gv} > 2 H$. 

Following \cite{Enqvist:2012tc} and the analytical method of estimating the number density of producing particles in the expanding universe presented there for which
\begin{equation}
 n^{(j)}_\chi \sim n^{(1)}_\chi \cdot 3^{j-1} \left( \frac{5}{2} \right)^{3/2}\frac{1}{j^{5/2}},
\end{equation}
where $j$ denotes the number of oscillations, we can see the agreement with our results. If we take $j\sim10$ as in the Figure \ref{fig:two} and $n_\chi^{(10)}\sim1\times10^{-6}$, we can see that the oscillation phase indeed finishes when $\frac{1}{2}m_\phi^2\langle\phi_j\rangle^2\sim \rho_\chi^{(j)}\sim g\langle\phi_j\rangle n_\chi^{(j)}$. 

The distribution of the produced states is not thermal but, assuming that the whole energy is transferred to the light states which interact with each other and with other particles not present in the simplified Lagrangian, we can naively estimate the maximal reheating temperature as
 \begin{equation}
T_{R}^{\text{max}} \sim \Big( \frac{30 \rho_R}{g_* \pi^2} \Big)^{1/4},
 \end{equation}
 where $\rho_R$ is energy density of the relativistic particles (in our case $\chi$ or $\chi$ and $\xi$) and $g_*$ describes the number of relativistic degrees of freedom ($g_* \sim \mathcal{O}(10^2)$). In our system the coupling is big enough to describe energy density as $\rho = m n$ and without contradicting our assumptions we can choose the masses as in Table \ref{tab:1}. Final estimation of $T_{R}^{\text{max}}$ is also presented in Table \ref{tab:1}.

\begin{table}[t]
\caption{Energy densities and upper limits on reheating temperature for two choices of $\chi$ mass. Mass of $\phi$ is set to \mbox{$m_{\phi} = 5 \cdot 10^{14} \text{ GeV}$}. Number densities for each state correspond to the results from Figure \ref{fig:two}, meaning that \mbox{$n_{\phi} \approx 3.96 \cdot 10^{-2}$ GeV$^3$} and $n_{\chi} \approx 8.2 \cdot 10^{-9}$ GeV$^3$.}
\label{tab:1}
\begin{tabular}{ccc}
\hline  \hline
    $m_{\chi}$ [GeV] & $\rho_{\chi}$ [GeV$^4$] & $T_R^{\text{max}}$ [GeV] \\
  \hline  
   &  &   \\
    125 & $ 10^{-6}$ & $1.3 \cdot 10^{-2}$ \\
  700 & $5.7 \cdot 10^{-6}$ & $2 \cdot 10^{-2}$ \\
   &  &  \\
  \hline  \hline
  \end{tabular}
\end{table}

\subsection{System with the additional light sector}

Usually when describing preheating light fields not coupled directly to the inflaton are neglected. But it is important to note that corresponding particles may be produced through an interaction with some other state coupled directly to the background that is produced resonantly. Furthermore, if there are many additional light degrees of freedom, one can expect that energy transfer from the background to the light sector during preheating might be sizeable. In this section we focus on such light fields and discuss the possibility of their production through the indirect interaction with the background field.

We can describe such a situation by extending (\ref{lagr2}) with $n$ light or massless fields $\xi_n$ ($m_\phi \gg m_\chi, m_\xi$) that are not coupled to the background at the tree-level
\begin{eqnarray}
& \nonumber \mathcal{L} = \frac{1}{2}(\partial \phi)^2 + \frac{1}{2}(\partial \chi)^2 - \frac{1}{2}m_\phi^2 \phi^2 - \frac{1}{2}m_\chi^2 \chi^2 - \frac{1}{4}g^2\phi^2\chi^2 \\
& + \sum_n\frac{1}{2}(\partial \xi_n)^2 - \sum_n\frac{1}{2}m_\xi^2 \xi_n^2 - \sum_n\frac{1}{4}y^2\chi^2\xi_n^2.
\end{eqnarray}
We assume again that $\langle \phi \rangle$ is time-varying and the other fields do not have a vev: $ \langle \chi \rangle = \langle \xi_n \rangle = 0 $. Then $\chi$ particles are produced resonantly and as we mentioned before we can expect production of $\xi_n$ through the interactions with $\chi$.

The physical mass of $\xi_n$ is given by
\begin{eqnarray}
\nonumber & M^2_{\xi} = m_{\xi}^2 + \frac{1}{2} y^2 \int \frac{d^3 p}{(2 \pi)^3} \left( \frac{1}{V} \langle \hat{\chi}^{\dagger}_{\textbf{p}} \hat{\chi}_{\textbf{p}} \rangle - \frac{1}{2 \omega_{\chi p}} \right) \\
& + \mathcal{O} (y^4, y^2 g^2, g^4),
\end{eqnarray}
where $\omega_{\chi p} \equiv \sqrt{\textbf{p}^2 + M_{\chi}^2} $ and $V$ denotes the volume of the system. We can see that $\xi_n$ influence background's evolution via $\langle \chi^{\dagger}_{\textbf{p}} \chi_{\textbf{p}} \rangle$ operator in their mass term.

\begin{figure}[h!]
 \begin{center}
\scalebox{0.35}{\includegraphics[width=1.1\textwidth]{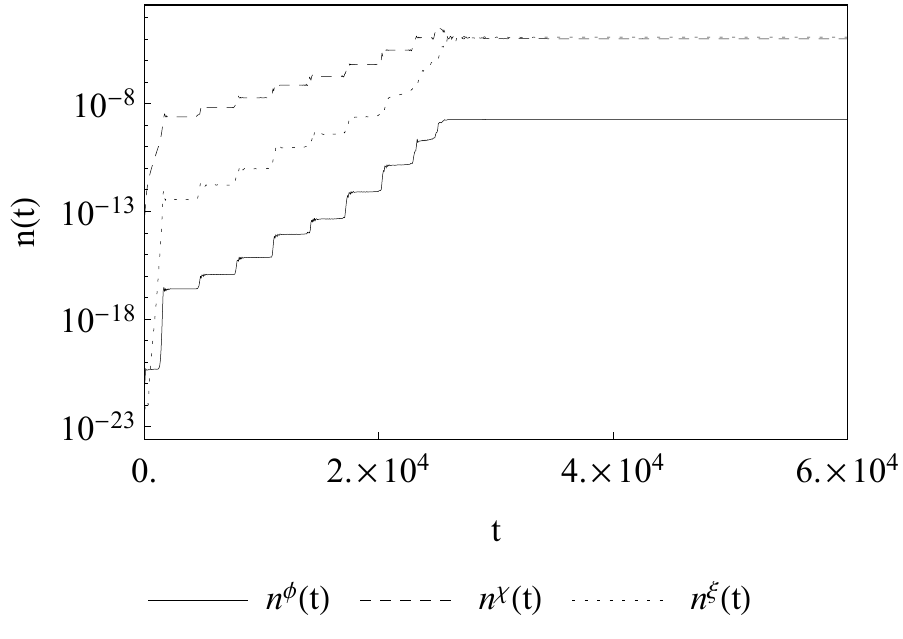}}
   \caption{Time evolution of number density of produced states in the system with additional light sector for $g=0.1$, $y=1$, $n=1$, $m_\phi=0.001 M$, $\phi(t=0)=M$, $\dot{\phi}(t=0)=0$. }
  \label{fig:quench}
  \end{center}
\end{figure}

We show the results for only one additional field in Figure \ref{fig:quench}. One can see that all the states are produced and their number density is abundant. If the final number density of $\xi$ is comparable to the one for $\chi$ ($n_{\xi} \sim n_{\chi}$) its presence may even quench the preheating process by terminating the energy transfer. The reason that $\xi$ can be produced so efficiently is the strong coupling between $\chi$ and $\xi$ that enhances the back-reaction effects.

\begin{figure}[h!]
 \begin{center}
 \scalebox{0.35}{ \includegraphics[width=1.1\textwidth]{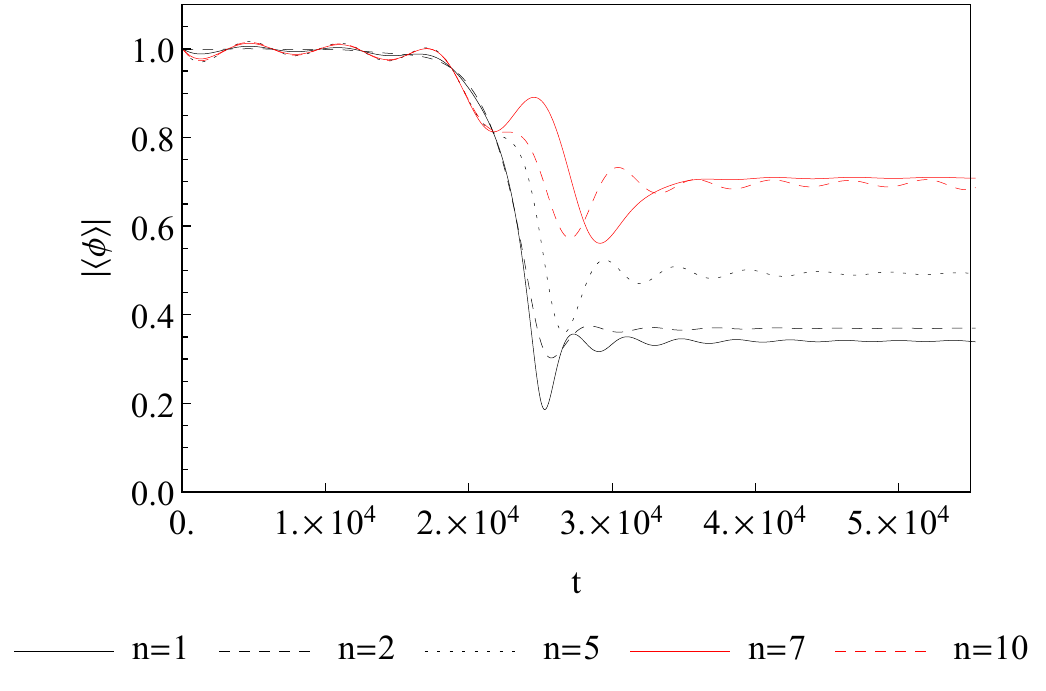}}
  \caption{Envelope of the time evolution of the background $\left<\phi\right>$ for $g=0.1$, $y=1$, $m_\phi=0.001 M$, $\phi(t=0)=M$, $\dot{\phi}(t=0)=0$ for different numbers of additional light fields: 1, 2, 5, 7.}
  \label{fig:quenchn}
  \end{center}
\end{figure}

We would expect that most of the energy would be transferred to $\xi_n$ fields as they are very light and the process is energetically favourable. But we can prove that the more light species we include, the larger the final value of $| \langle \phi \rangle |$ becomes and, in other words, the less energy from the background goes to the light fields, see Figure \ref{fig:quenchn}. 

The reason why the energy transfer can be stopped in this case can be understood as follows. The physical mass of $\chi$ in the system is given by
\begin{eqnarray}
& \nonumber M_{\chi}^2 = m_{\chi}^2 + \frac{1}{2} g^2 \langle \phi \rangle^2 + \frac{1}{2} g^2 \int \frac{d^3 p}{(2 \pi)^3} \left( \frac{1}{V} \langle \hat{\phi}_{\textbf{p}}^{\dagger} \hat{\phi}_{\textbf{p}} \rangle - \frac{1}{2 \omega_{\phi p}} \right) \\
& + \frac{1}{2} y^2 \sum \limits_n \left( \frac{1}{V} \langle \hat{\xi}_{n \textbf{p}}^{\dagger} \hat{\xi}_{n \textbf{p}} \rangle - \frac{1}{2 \omega_{\xi p}} \right) + \mathcal{O} (y^4, y^2 g^2, g^4).
\end{eqnarray}
Considering an approximation $\dot{\hat{X}}_{\textbf{p}} \sim -i \omega_{X p} \hat{X}_{\textbf{p}} $ for \mbox{$X = \phi, \xi_n $}, one can find that
\begin{eqnarray}
& \frac{1}{V} \langle \hat{X}^{\dagger}_{\textbf{p}} \hat{X}_{\textbf{p}} \rangle - \frac{1}{2 \omega_{X_p}} \sim \frac{1}{2 \omega_{X_p}} \frac{1}{V} \left \langle N^{(+)}_{X_\textbf{p}} \right \rangle.
\end{eqnarray}
Thus, once $\phi$ or $\xi_n$ are produced at the same time they also generate $\chi$'s effective mass \footnote{These mass correction terms describe a square of \textit{plasma frequency} discovered by I.Langmuir and L.Tonks in the 1920s which is a critical value for which the wave of $\chi$ can enter $X$'s plasma or not, because 
\begin{equation}
\nonumber \int\frac{d^3p}{(2\pi)^3}\frac{\langle N_{\mathbf{k}}^{(+)}\rangle}{2\omega_{Xp}V}
\end{equation}
is proportional to $\frac{n_X}{M_X}$ if $X$ particles are massive enough ($n_X$ is $X$'s number density). Moreover, if one considers the massless thermal equilibrium distribution with the temperature $T$: 
\begin{equation}
\nonumber \frac{1}{V} \langle N^{(+)}_{X \textbf{p}} \rangle = 2 \frac{1}{e^{p/T} - 1}
\end{equation}
(factor 2 corresponds to the degrees of freedom for momentum $\mathbf{k}$ and $\mathbf{-k}$ particles), it corresponds to the thermal mass of the form: 
\begin{equation}
\nonumber \int \frac{d^3 p}{(2 \pi)^3} \left( \frac{1}{V} \langle \hat{X}_{\textbf{p}}^{\dagger} \hat{X}_{\textbf{p}} \rangle - \frac{1}{2 \omega_{X_p}} \right) \sim \frac{T^2}{6}.
\end{equation}} which results in $\chi$ particle production area becoming narrower. This leads to the suppression of $\chi$ particle production and also spoils the production of other species. Too many $\xi_n$ particles produced through indirect coupling to the background prevent the production of particles directly coupled to the background, $\chi$. 

\begin{figure}[h!]
 \begin{center}
   \scalebox{0.35}{ \includegraphics[width=1.1\textwidth]{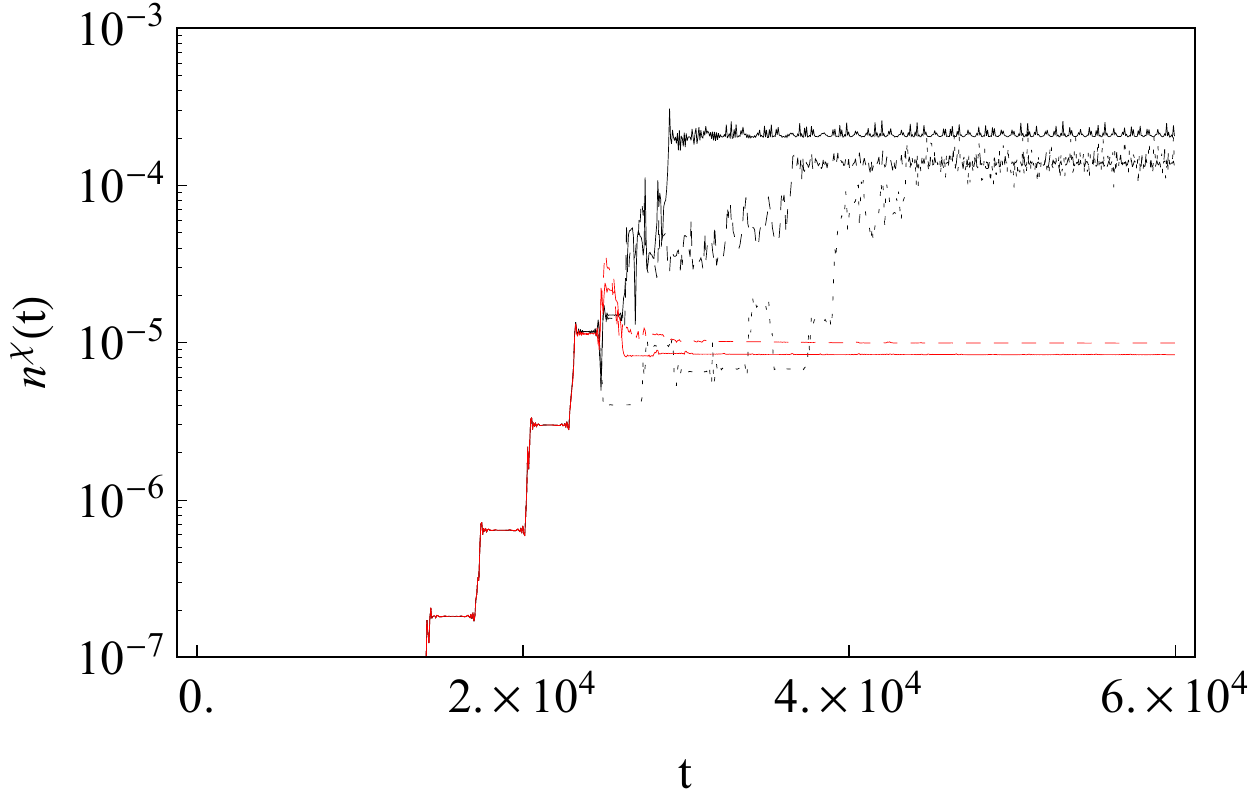}}
    \scalebox{0.35}{ \includegraphics[width=1.1\textwidth]{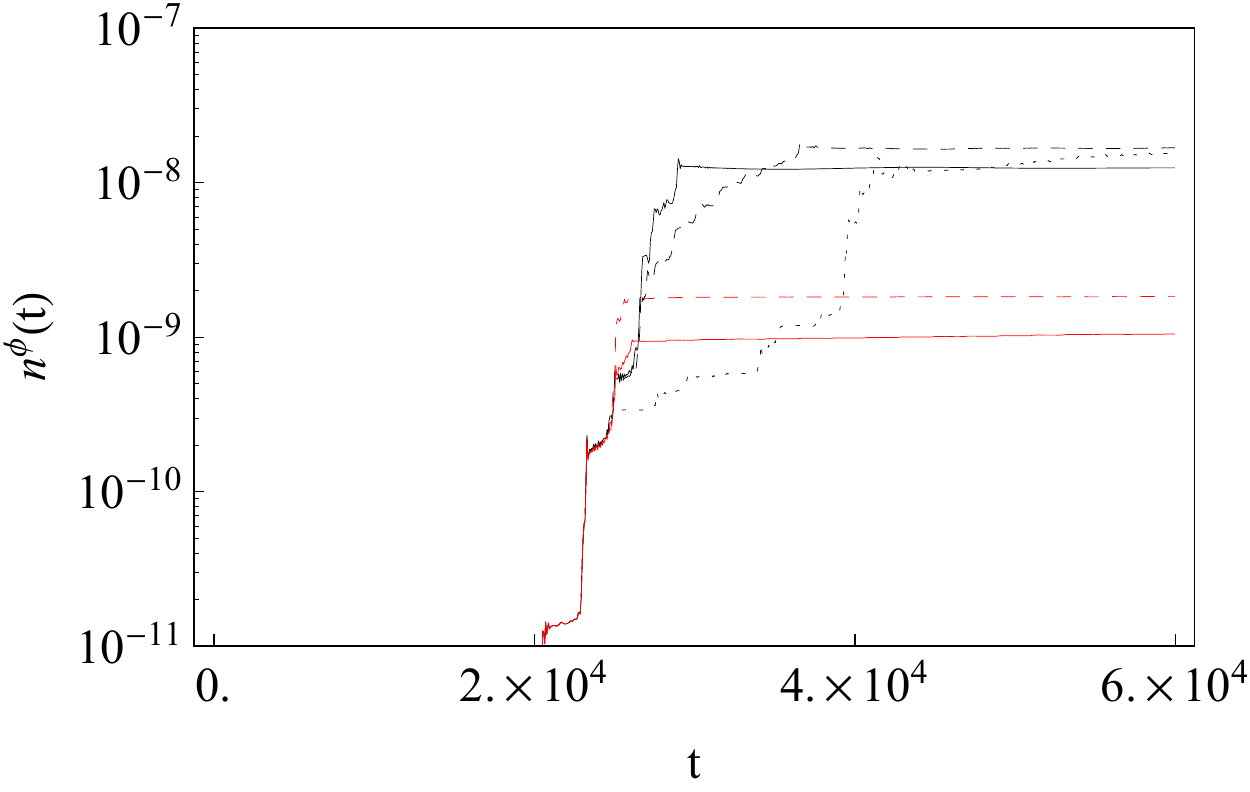}}
      \scalebox{0.35}{   \includegraphics[width=1.1\textwidth]{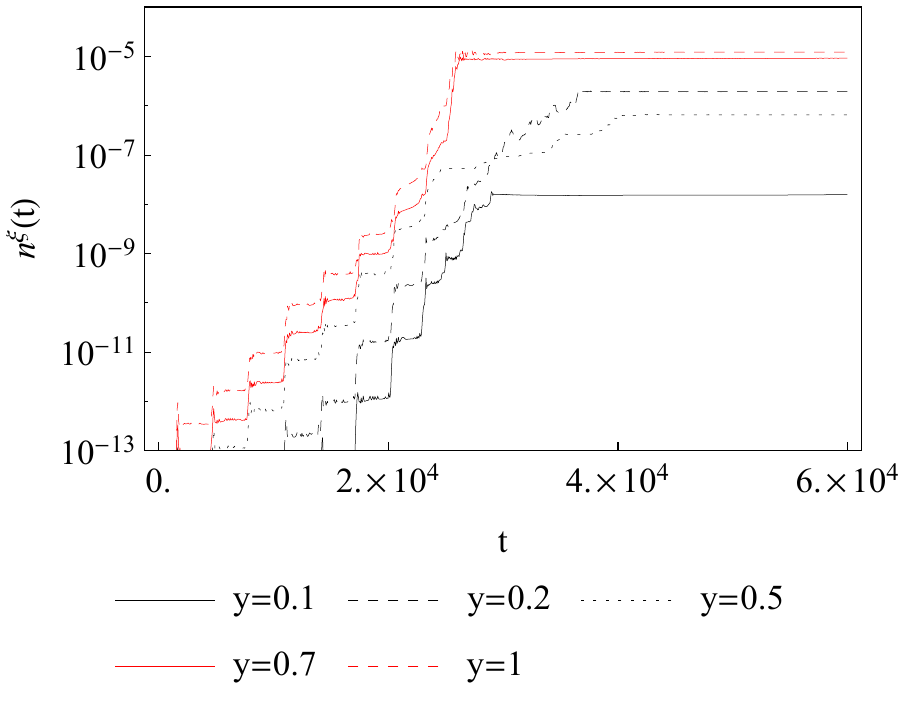}}
  \caption{Time evolution of number density of produced states $\chi$, $\phi$ and $\xi$ for $g=0.1$, $n=1$, $m_\phi=0.001 M$, $\phi(t=0)=M$, $\dot{\phi}(t=0)=0$ and different values of $y$ coupling. Values $y=0.7$ and $y=1$ correspond to quenching of parametric resonance.}
   \label{fig:param-dens}
    \end{center}
\end{figure}

\begin{figure}[h!]
 \begin{center}
  \scalebox{0.35}{  \includegraphics[width=1.2\textwidth]{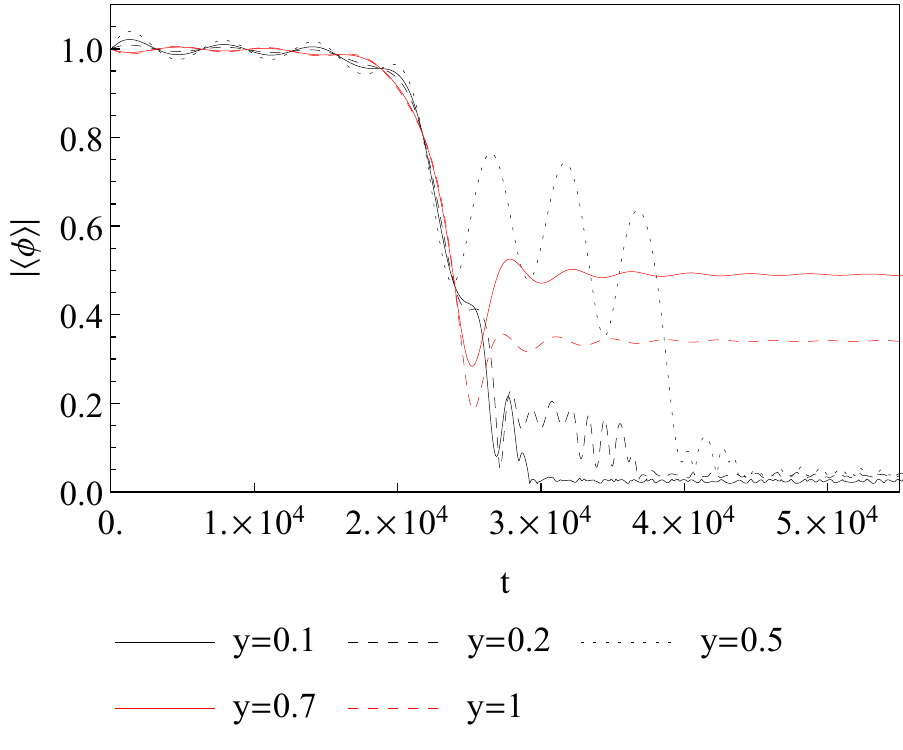}}
  \caption{Envelope of the time evolution the background $\left<\phi\right>$ for $g=0.1$, $n=1$, $m_\phi=0.001 M$, $\phi(t=0)=M$, $\dot{\phi}(t=0)=0$ and different values of $y$ coupling. For $ y=0.7$ and $y=1$ we can observe the quenching of the preheating.}
   \label{fig:param-back}
    \end{center}
\end{figure}

It is interesting to investigate the impact of both couplings - $g$ that couples $\chi$ to $\phi$ and the background $\langle \phi \rangle$ and $y$ that couples additional fields $\xi_n$ to $\chi$, on the features of preheating. Varying the coupling $y$ for fixed $g$ leads to the conclusion that the initial stage of preheating does not depend on $y$ coupling for $\chi$ and $\phi$ states. It only influences the final abundance of produced $\chi$ and $\phi$ states - the bigger $y$ is, the smaller number density of these states we observe, see Figure \ref{fig:param-dens}. For $\xi$ the impact of $y$ is quite opposite - both initial and final stages of $\xi$ production are strongly influenced by the value of $y$. This time the bigger $y$ is, the larger number density of $\xi$ we observe which also results in more effective energy transfer to the background as $y$ coupling drops, see Figure \ref{fig:param-back}. Also, for choices of parameters resulting in $n_{\xi} \sim n_{\chi}$ we can observe quenching of the energy transfer from the background.

Our study may seem similar to the process of instant preheating \citep{Felder:1998vq, Tsujikawa:2000ik}, where the system of three fields - background $\phi$, $\chi$ interacting with the background and some other field $\psi$ not coupled to $\phi$, is considered. Instant preheating relies on the fact that $\chi$ particles produced within one-time oscillation of $\phi$ decay immediately to $\psi$ before the next oscillation of $\phi$. So $\psi$ states can be also produced even though there is no direct interaction between $\phi$ and $\psi$. In our work the mechanism of production is different - due to the quantum corrections, not the decay, and quenching of the preheating comes from a plasma gas effect here rather than the rapid decay.

Table \ref{tab:2} presents $T_R^{\text{max}}$ and energy densities for each state for the considered model under assumption that $\chi = H$ or $\xi = H$, $H$ being the Higgs field playing the role of the mediator or the light field. We can see that additional light sector that quenches preheating rises $T_R$ lowering the number density of $\phi$ particles at the same time.

\begin{table}[t]
\caption{Energy densities and upper limits on reheating temperature (both in GeV) for two choices of $\chi$ and $\xi$ mass. Mass of $\phi$ is set to \mbox{$m_{\phi} = 5 \cdot 10^{14} \text{ GeV}$}. Number densities for each state correspond to the results from Figure \ref{fig:quench}, meaning that $n_{\phi} \approx 1.82 \cdot 10^{-9}$ GeV$^3$ and $n_{\chi} \approx 9.91 \cdot 10^{-6}$ GeV$^3$.}
\label{tab:2}
\begin{tabular}{cccccc}
\hline \hline
     $m_{\chi}$ [GeV] & $m_{\xi}$ [GeV]  & $n_{\xi}$ [GeV$^3$] & $\rho_{\chi}$ [GeV$^4$] & $\rho_{\xi}$ [GeV$^4$] & $T_R^{\text{max}}$ [GeV] \\
\hline  
&  &   &  &  &  \\
    125 & 100 & $1.21 \cdot 10^{-5}$ & $1.24 \cdot 10^{-3}$ & $1.21 \cdot 10^{-3}$ & $0.93 \cdot 10^{-1}$ \\
  700 & 125 & $1.21 \cdot 10^{-5}$ & $6.94 \cdot 10^{-3}$ & $1.51 \cdot 10^{-3}$ & $1.26 \cdot 10^{-1}$ \\
 &  &  &   &  &  \\ 
  \hline \hline
    \end{tabular}
\end{table}

\section{Discussion and summary \label{sec_disc}}
	
\begin{figure}[h!]
 \begin{center}
\scalebox{0.35}{ \includegraphics[width=1.1\textwidth]{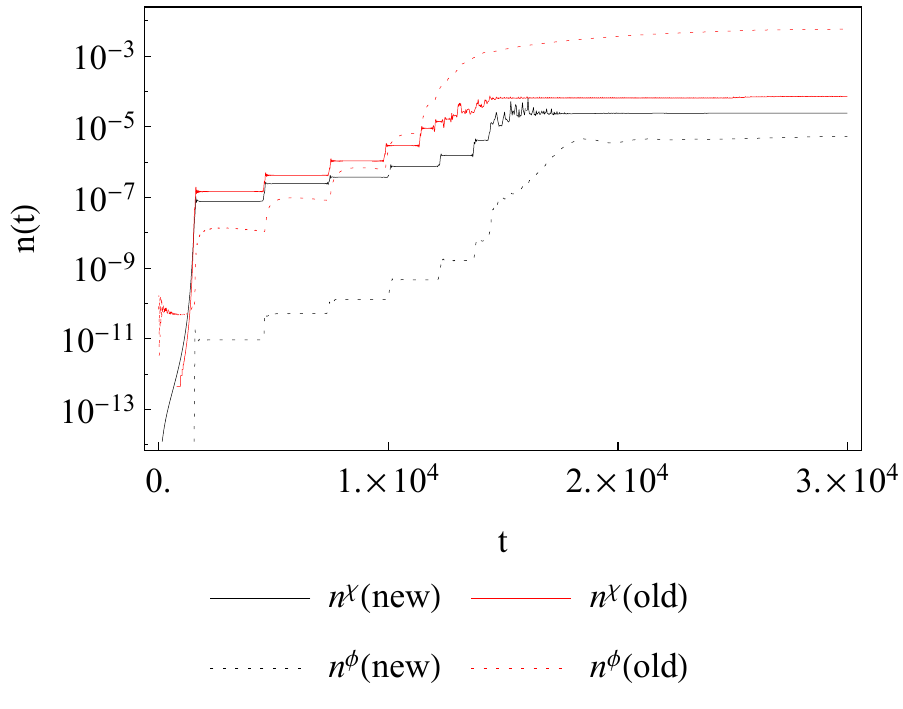}}\\
\scalebox{0.35}{    \includegraphics[width=1.1\textwidth]{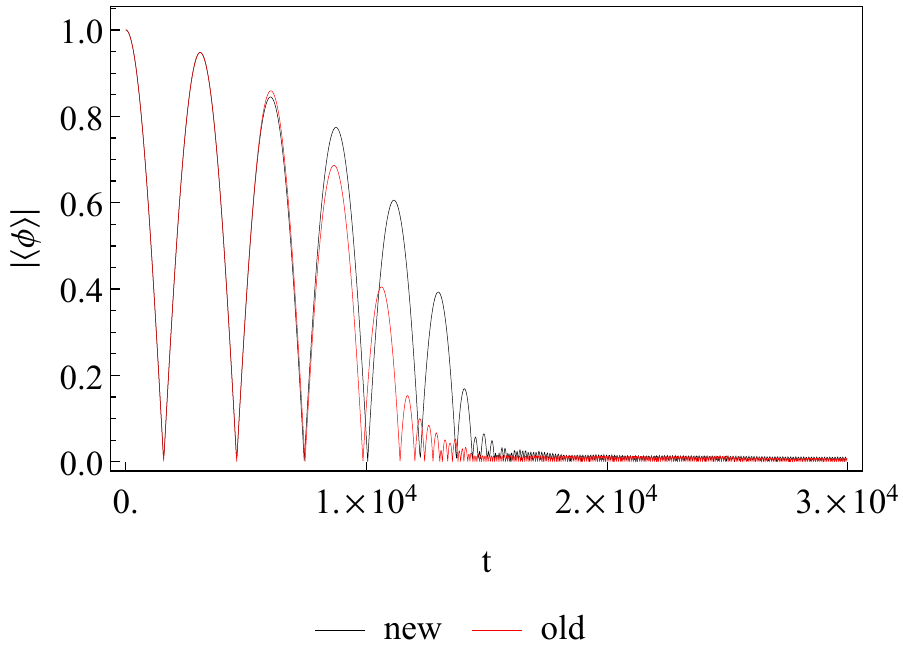}}
  \caption{Comparison between time evolution of number density of produced states (\textit{upper}) and the background $\left<\phi\right>$ (\textit{lower}) obtained with a \textit{new} and \textit{old} methods for $g=1$, $m_\phi=0.001 M$, $\phi(t=0)=M$, $\dot{\phi}(t=0)=0$. \textit{New} denotes the interacting theory described here and \textit{old} - asymptotic approximation presented in \citep{Enomoto:2014cna}.}
   \label{fig:old_vs_new}
    \end{center}
\end{figure}

In our previous work \cite{Enomoto:2014cna} we presented a formalism for describing particle production in a time-dependent background. It turned out it possesses one drawback - there exists a secularity in the number density of massless states that can be a product of approximating the fields by their asymptotic values. In this paper we have developed more accurate description by expressing the number operator in terms of interacting fields. Figure \ref{fig:old_vs_new} compares the two methods for the Lagrangian (\ref{lagr2}). The new method avoids artificial secularity caused by time integral of the interaction effects with the Green functions seen before. The old method seems to overestimate the production at the late stage because it includes "inverse decay" processes, whereas the new one takes into account mass correction terms. However, the results with secularity are still applicable at the early stages of particle production process. 

As the application of the new method in this paper we investigated the role of additional light fields coupled indirectly to the background during resonant particle production processes such as preheating. In particular, we considered models with a scalar field $\chi$ interacting with the background $\langle \phi \rangle$ through its mass term and with $n$ light fields $\xi_n$. In order to describe particle production in the system, at first we defined number operator in terms of interacting fields and then we solved numerically their equations of motion. In case of a few additional light fields, their production can be also resonant through the quantum correction to their mass term and their final amount can be sizeable. However, many degrees of freedom of these extra light fields can prevent $\chi$'s and also $\xi_n$'s resonant particle production. As a result, energy transfer from the background does not work well and this indicates that preheating might be quenched if there are many degrees of freedom of light fields which are connected to the background indirectly.\\

This work has been supported by the Polish NCN grant DEC-2012/04/A/ST2/00099, OC was also supported by the doctoral scholarship number 2016/20/T/ST2/00175. SE is partially supported by the Heising-Simons Fundation grant No 2015-109. OC thanks Bonn Bethe Centre Theory Group for hospitality during the completion of this paper.

\section*{Appendix A: Particle production in free fields theory with time-varying mass terms}

Let us consider a real free scalar field $\phi$ with the time-dependent mass term:
\begin{equation}
 \mathcal{L} = \frac{1}{2} (\partial \phi)^2 - \frac{1}{2}m^2(t) \phi^2.
\end{equation}
The solution of the equation of motion can be decomposed into 
\begin{equation}
 \phi(x) = \int \frac{d^3k}{(2\pi)^3} e^{i\mathbf{k \cdot x}}
  \left( \phi_k a_{\mathbf{k}} + \phi_k^* a_{-\mathbf{k}}^{\dagger} \right)  \label{eq:solution_time_dep}
\end{equation}
where $\phi_k=\phi_k(x^0)$ is a time-dependent wave function which satisfies
\begin{equation}
 0 = \ddot{\phi}_k + \omega_k^2 \phi_k \quad ( \omega_k\equiv\sqrt{k^2+m^2}),\label{eq:eom_wave_time_dep}
\end{equation}
and $a_{\mathbf{k}}, a_{\mathbf{k}}^\dagger$ are annihilation and creation operators. The vacuum state $|0\rangle$ is defined by the relation $a_{\mathbf{k}}|0\rangle=0$ and the commutation relations
\begin{eqnarray}
 &[\phi(t,\mathbf{x}),\dot{\phi}(t,\mathbf{x'})]=i\delta(\mathbf{x-x'}),&\\
 &[\phi(t,\mathbf{x}),\phi(t,\mathbf{x'})]=[\dot{\phi}(t,\mathbf{x}),\dot{\phi}(t,\mathbf{x'})]=0,&\\
 &[a_{\mathbf{k}}, a_{\mathbf{k'}}^\dagger]=(2\pi)^3\delta(\mathbf{k-k'}), &\\ & [a_{\mathbf{k}}, a_{\mathbf{k'}}]=[a_{\mathbf{k}}^\dagger, a_{\mathbf{k'}}^\dagger]=0&
\end{eqnarray}
give an inner product relation of the form: $(\phi_k,\phi_k)=1$.

Using (\ref{eq:solution_time_dep}) we can represent the Hamiltonian as
\begin{eqnarray}
 H &=& \int d^3x \left( \frac{1}{2}\dot{\phi}^2 + \frac{1}{2}(\nabla \phi)^2 + \frac{1}{2}m^2\phi^2 \right) \\
  &=& \int \frac{d^3k}{(2\pi)^3} \frac{1}{2} \left[ \Omega_k(t) \left( a_{\mathbf{k}}^{\dagger} a_{\mathbf{k}} + a_{\mathbf{-k}} a_{\mathbf{-k}}^{\dagger} \right) \right. \nonumber \\
  & & \qquad \qquad \left. + \Lambda_k(t) a_{\mathbf{-k}} a_{\mathbf{k}} + \Lambda_k^*(t) a_{\mathbf{k}}^{\dagger} a_{\mathbf{-k}}^{\dagger} \right],
\end{eqnarray}
where
\begin{eqnarray}
 \Omega_k(t) & \equiv & |\dot{\phi}_k(t)|^2 + \omega_k^2(t) |\phi_k(t)|^2, \\
 \Lambda_k(t) & \equiv & \dot{\phi}_k^2(t) + \omega_k^2(t) \phi_k^2(t).
\end{eqnarray}
In order to diagonalize the Hamiltonian
\begin{eqnarray}
 H &=& \int \frac{d^3k}{(2\pi)^3} \frac{1}{2}\omega_k(t) \left( \bar{a}_{\mathbf{k}}^{\dagger} \bar{a}_{\mathbf{k}} + \bar{a}_{\mathbf{-k}} \bar{a}_{\mathbf{-k}}^{\dagger} \right) \\
  &=&  \int \frac{d^3k}{(2\pi)^3} \omega_k(t) \left( \bar{a}_{\mathbf{k}}^{\dagger} \bar{a}_{\mathbf{k}} + \frac{1}{2} (2\pi)^3\delta^3(\mathbf{k}=0)\right)
\end{eqnarray}
we need a set of operators $\bar{a}_{\mathbf{k}}, \bar{a}_{\mathbf{k}}$ satisfying \begin{equation}
 [\bar{a}_{\mathbf{k}}, \bar{a}^{\dagger}_{\mathbf{k'}}]=(2\pi)^3\delta(\mathbf{k-k'}),\:[\bar{a}_{\mathbf{k}}, \bar{a}_{\mathbf{k'}}]=[\bar{a}_{\mathbf{k}}^\dagger, \bar{a}_{\mathbf{k'}}^\dagger]=0,
\end{equation}
Then the number operator $\bar{N}_\mathbf{k}\equiv \bar{a}_{\mathbf{k}}^\dagger \bar{a}_{\mathbf{k}}$ is well-defined all the time. Following \cite{Garbrecht:2002pd}, we can obtain the new operators by the Bogoliubov transformation
\begin{equation}
 \bar{a}_{\mathbf{k}}=\alpha_ka_{\mathbf{k}}+\beta_ka_{\mathbf{-k}}^\dagger
\end{equation}
with the coefficients satisfying
\begin{equation}
 |\alpha_k|^2=\frac{\Omega_k^{\rm in}}{2\omega_k}+\frac{1}{2}, \:
 |\beta_k|^2=\frac{\Omega_k^{\rm in}}{2\omega_k}-\frac{1}{2}, \:
 {\rm Arg}(\alpha_k \beta_k^*) = {\rm Arg} \: \Lambda_k^{\rm in}.
\end{equation}
Then the occupation number can be expressed as
\begin{equation}
 N_k(t)=\langle 0|\bar{N}_\mathbf{k}|0\rangle=|\beta_k|^2=\frac{\Omega_k}{2\omega_k}-\frac{1}{2}.
\end{equation}

\section*{Appendix B: Two scalar system \\ - details of the calculation}

In the system described by the Lagrangian \ref{lagr}, we have a background field $\langle\phi\rangle$ and two quantum fields: $\tilde{\phi}\equiv\phi-\langle\phi\rangle$, $\chi$. The set of differential equations for distributions reads  
\begin{eqnarray}
 & 0  = \langle \ddot{\phi}\rangle+M_\phi^2\langle\phi\rangle\label{eq:eom_phi_2scalar}\\
& \langle \hat{\phi}_{\textbf{k}}^{\dagger} \hat{\phi}_{\textbf{k}} \rangle^{\cdot}
  = \langle \dot{\hat{\phi}}_{\textbf{k}}^{\dagger}\hat{\phi}_{\textbf{k}} \rangle + \langle \hat{\phi}_{\textbf{k}}^{\dagger} \dot{\hat{\phi}}_{\textbf{k}} \rangle \\
& \langle \hat{\phi}_{\textbf{k}}^{\dagger} \dot{\hat{\phi}}_{\textbf{k}} \rangle^{\cdot}
  = \langle \dot{\hat{\phi}}_{\textbf{k}}^{\dagger} \dot{\hat{\phi}}_{\textbf{k}} \rangle - \omega_{\phi k}^2 \langle \hat{\phi}_{\textbf{k}}^{\dagger} \hat{\phi}_{\textbf{k}} \rangle \\
& \langle \dot{\hat{\phi}}_{\textbf{k}}^{\dagger} \dot{\hat{\phi}}_{\textbf{k}} \rangle^{\cdot}
  = -\omega_{\phi k}^2(\langle \dot{\hat{\phi}}_{\textbf{k}}^{\dagger} \hat{\phi}_{\textbf{k}} \rangle + \langle \hat{\phi}_{\textbf{k}}^{\dagger}\dot{\hat{\phi}}_{\textbf{k}} \rangle)\\
& \langle \hat{\chi}_{\textbf{k}}^{\dagger} \hat{\chi}_{\textbf{k}} \rangle^{\cdot}
  = \langle \dot{\hat{\chi}}_{\textbf{k}}^{\dagger}\hat{\chi}_{\textbf{k}} \rangle + \langle \hat{\chi}_{\textbf{k}}^{\dagger} \dot{\hat{\chi}}_{\textbf{k}} \rangle \\
& \langle \hat{\chi}_{\textbf{k}}^{\dagger} \dot{\hat{\chi}}_{\textbf{k}} \rangle^{\cdot}
  = \langle \dot{\hat{\chi}}_{\textbf{k}}^{\dagger} \dot{\hat{\chi}}_{\textbf{k}} \rangle - \omega_{\chi k}^2 \langle \hat{\chi}_{\textbf{k}}^{\dagger} \hat{\chi}_{\textbf{k}} \rangle \\
& \langle \dot{\hat{\chi}}_{\textbf{k}}^{\dagger} \dot{\hat{\chi}}_{\textbf{k}} \rangle^{\cdot}
  = -\omega_{\chi k}^2(\langle \dot{\hat{\chi}}_{\textbf{k}}^{\dagger} \hat{\chi}_{\textbf{k}} \rangle + \langle \hat{\chi}_{\textbf{k}}^{\dagger}\dot{\hat{\chi}}_{\textbf{k}} \rangle)\label{eq:eom_dcdc_d_2scalar}
\end{eqnarray}
where the source terms are absent because of our choice of physical masses
\begin{eqnarray}
  M_\phi^2 &=& m_\phi^2+\frac{1}{2}g^2\int\frac{d^3p}{(2\pi)^3}\left[\frac{1}{V}\langle\hat{\chi}_{\textbf{p}}^{\dagger}\hat{\chi}_{\textbf{p}} \rangle -\frac{1}{2\omega_{\chi p}}\right],\\
  M_\chi^2 &=& m_\phi^2+\frac{1}{2}g^2\langle\phi\rangle^2\nonumber\\
  & & \quad+\frac{1}{2}g^2\int\frac{d^3p}{(2\pi)^3}\left[\frac{1}{V}\langle\hat{\phi}_{\textbf{p}}^{\dagger}\hat{\phi}_{\textbf{p}} \rangle -\frac{1}{2\omega_{\phi p}}\right].
\end{eqnarray}
In order to obtain the above formulae, we applied an approximation
\begin{equation}
 \langle{\hat{\phi}}_{\mathbf{p_1}}^\dagger\hat{\phi}_{\mathbf{p_2}}\hat{\chi}_{\mathbf{p_3}}^\dagger\hat{\chi}_{\mathbf{p_4}}\rangle=
 \langle{\hat{\phi}}_{\mathbf{p_1}}^\dagger\hat{\phi}_{\mathbf{p_2}}\rangle\langle\hat{\chi}_{\mathbf{p_3}}^\dagger\hat{\chi}_{\mathbf{p_4}}\rangle+\mathcal{O}(g^2)
\end{equation}
and assumed the momentum conservation
\begin{equation}
 \langle{\hat{X}}_{\mathbf{p}}^\dagger\hat{X}_{\mathbf{p'}}\rangle=\frac{1}{V}(2\pi)^3\delta^3(\mathbf{p-p'})\cdot\langle{\hat{X}}_{\mathbf{p}}^\dagger\hat{X}_{\mathbf{p}}\rangle
\end{equation}
for the quantum fields $X=\tilde{\phi},\chi$. Momentum conservation indicates that $\langle{\hat{X}}_{\mathbf{p}}^\dagger\hat{X}_{\mathbf{p'}}\rangle= C_{\mathbf{p}} (2\pi)^3\delta^3(\mathbf{p-p'})$, where $C_{\mathbf{p}}$ is a proportionality factor. For $\mathbf{p'}=\mathbf{p}$: $\langle{\hat{X}}_{\mathbf{p}}^\dagger\hat{X}_{\mathbf{p}}\rangle=V\cdot C_{\mathbf{p}}$, hence $C_{\mathbf{p}}=\frac{1}{V}\langle{\hat{X}}_{\mathbf{p}}^\dagger\hat{X}_{\mathbf{p}}\rangle$.


\begin{thebibliography}{99}

\bibitem{Kofman:1994rk}
      L.~Kofman, A.~Linde, A.~Starobinsky,
      Phys. Rev. Lett. \textbf{73} (1994)
      \mbox{[hep-th/9405187]}.
      
\bibitem{Kofman:1997yn}
      L.~Kofman, A.~Linde, A.~Starobinsky,
      Phys. Rev. D \textbf{56} (1997)
      \mbox{[hep-ph/9704452]}.
      
\bibitem{Traschen:1990sw}
  J.~H.~Traschen and R.~H.~Brandenberger,
  Phys.\ Rev.\ D {\bf 42} (1990) 2491.
  
\bibitem{Dolgov:1989us}
  A.~D.~Dolgov and D.~P.~Kirilova,
  Sov.\ J.\ Nucl.\ Phys.\  {\bf 51} (1990) 172
   [Yad.\ Fiz.\  {\bf 51} (1990) 273].
      
\bibitem{Kofman:2004yc}
      L.~Kofman \textit{et al.},
      JHEP \textbf{05} (2004)
      \mbox{hep-th/0403001}.

\bibitem{Kobayashi:2010fm}
      T.~Kobayashi, S.~Mukohyama, 
      Phys. Rev. D \textbf{81} (2010)
      \mbox{[astro-ph.CO/1003.0076]}.
      
\bibitem{Matsuda:2012kk}
      T.~Matsuda, 
      JCAP \textbf{1204} (2012)
      [hep-ph/1204.0303].      

\bibitem{Kohri:2014jma}
      K.~Kohri, T.~Matsuda, 
      JCAP \textbf{1502} (2015)
      \mbox{[astro-ph.CO/1405.6769]}. 
     
\bibitem{Enqvist:2001zp}
      K.~Enqvist, M.~Sloth, 
      Nucl. Phys. B \textbf{626} (2002)
      \mbox{[hep-ph/0109214]}.

\bibitem{Lyth:2001nq}
      D.~Lyth, D.~Wands, 
      Phys. Lett. B \textbf{524} (2002)
      \mbox{[hep-ph/0110002]}.

\bibitem{Moroi:2001ct}
      T.~Moroi, T.~Takahashi, 
      Phys. Lett. B \textbf{522} (2001)
      \mbox{[hep-ph/0110096]}. 
      
\bibitem{Allahverdi:2010xz}
  	  R.~Allahverdi, R.~Brandenberger, F.~Y.~Cyr-Racine and A.~Mazumdar,
      Ann.\ Rev.\ Nucl.\ Part.\ Sci.\  {\bf 60} (2010) 27
      [hep-th/1001.2600].
      
\bibitem{Amin:2014eta}
	  M.~A.~Amin, M.~P.~Hertzberg, D.~I.~Kaiser and J.~Karouby,
  	  Int.\ J.\ Mod.\ Phys.\ D {\bf 24} (2014) 1530003
  	  [hep-ph/1410.3808].
 
\bibitem{Enomoto:2014cna}
      S.~Enomoto, O.~Fuksi{\'n}ska, Z.~Lalak, 
      JHEP \textbf{03} (2015)
      [hep-ph/1412.7442].
       

\bibitem{Enqvist:2012tc} 
  K.~Enqvist, D.~G.~Figueroa and R.~N.~Lerner,
  JCAP {\bf 1301}, 040 (2013)
  [astro-ph.CO/1211.5028 ].
      
\bibitem{Felder:1998vq}
	G. Felder, L. Kofman, and A. Linde,
	Phys Rev D {\bf 59} (1999) 123523
	[hep-ph/9812289].

\bibitem{Tsujikawa:2000ik}
	S. Tsujikawa, B. Bassett, and F. Viniegra 
	JHEP {\bf 19} (2000)
	[hep-ph/0006354].


\bibitem{Garbrecht:2002pd} 
  B.~Garbrecht, T.~Prokopec and M.~G.~Schmidt,
  Eur.\ Phys.\ J.\ C {\bf 38}, 135 (2004)
  [hep-th/0211219].
\end{thebibliography}
\end{document}